\documentclass[10pt,twocolumn,twoside]{IEEEtran}

\usepackage{color}
\usepackage{CJK,indentfirst}
\usepackage[compress]{cite}
\makeatletter
\renewcommand{\citepunct}{,\penalty\@m\hskip.13emplus.1emminus.1em}
\renewcommand{\citedash}{\hbox{--}\penalty\@m}
\makeatother
\usepackage{graphicx}
\usepackage{amsmath}
\usepackage{amssymb}
\usepackage{amsfonts}
\usepackage{booktabs}
\usepackage{array}
\ifCLASSOPTIONcompsoc
\usepackage[tight,normalsize,sf,SF]{subfigure}
\else
\usepackage[tight,footnotesize]{subfigure}
\fi

\usepackage{threeparttable}
\usepackage{nccmath}

\newcommand{\bs}{\text{BS}}
\newcommand{\ms}{\text{MS}}

\begin{document}


\title{User Scheduling for Cooperative Base Station Transmission
Exploiting Channel Asymmetry}

\author{\baselineskip=0.75\normalbaselineskip
Shengqian Han, \IEEEmembership{Member, IEEE,} Chenyang Yang, \IEEEmembership{Senior Member, IEEE,} \\
and Mats Bengtsson, \IEEEmembership{Senior Member, IEEE}
\thanks{This work was supported in part by the International S\&T Cooperation Program
of China (ISCP) (No. 2008DFA12100) and the national key project of
next generation wideband wireless communication networks (No.
2011ZX03003-001-01).

S. Han and C. Yang are with the School of Electronics and
Information Engineering, Beihang University (BUAA), Beijing, China
(e-mail: sqhan@ee.buaa.edu.cn; cyyang@buaa.edu.cn). M. Bengtsson is
with the School of Electrical Engineering, Royal Institute of
Technology (KTH), Stockholm, Sweden (e-mail:
mats.bengtsson@ee.kth.se). }}

\maketitle
\begin{abstract}
We study low-signalling overhead scheduling for downlink coordinated
multi-point (CoMP) transmission with multi-antenna base stations
(BSs) and single-antenna users. By exploiting the asymmetric channel
feature, i.e., the pathloss differences towards different BSs, we
derive a metric to judge orthogonality among users only using their
average channel gains, based on which we propose a semi-orthogonal
scheduler that can be applied in a two-stage transmission strategy.
Simulation results demonstrate that the proposed scheduler performs
close to the semi-orthogonal scheduler with full channel
information, especially when each BS is with more antennas and the
cell-edge region is large. Compared with other overhead reduction
strategies, the proposed scheduler requires much less training
overhead to achieve the same cell-average data rate.
\end{abstract}

\section{Introduction}
Base station (BS) cooperative transmission, also known as
coordinated multi-point (CoMP) transmission, has received much
attention for providing high spectral efficiency in cellular
networks~\cite{Gesbert2010a}. By sharing data and channel state
information (CSI) among multiple BSs, CoMP-JP (joint processing) can
fully exploit the benefits of the cooperation. Among various
challenges such as BS synchronization, backhaul cost and channel
acquisition, the overhead to gather CSI is the most limiting factor
that hinders the application of CoMP.

Multi-user multi-input multi-output (MU-MIMO) techniques can exploit
the abundant spatial resources in downlink CoMP-JP systems. When the
number of users exceeds that of transmit antennas, spatial user
scheduling becomes critical~\cite{Yoo06-SUS}, which however requires
enormous training or feedback overhead even in single-cell
systems~\cite{Shi2010}. To reduce the overhead in case it
counteracts the performance gain, the training symbol length was
optimized to maximize a net throughput excluding the
overhead~\cite{6Hoydis2011}. The overhead can also be reduced by
differentiating what the CSI is used for. For example, when
zero-forcing beamforming (ZFBF) is employed, channel direction
information is essential for beamforming, and channel norms and
channel orthogonality among users are essential for a well known
semi-orthogonal user scheduler (SUS)~\cite{Yoo06-SUS}.
Alternatively, the overhead can be reduced by selective
feedback~\cite{Papadogiannis2011} or by exploiting channel
statistics~\cite{Hammarwall08}.

Though CoMP-JP exhibits many similarities to single-cell MU-MIMO,
there are distinctive differences in the channel properties and
system setting. One of them is the inherent \emph{channel asymmetry}
of CoMP systems, i.e., the average channel gains from multiple BSs
to each user are non-identical~\cite{Heath10ICASSP}. {Despite that
many well-explored precoders and schedulers can be directly applied
for CoMP-JP systems if full CSI of all candidate users is available,
such a unique channel feature can be exploited to further reduce the
required overhead.}

This paper aims at designing low-overhead scheduling for downlink
CoMP-JP systems. Considering that SUS~\cite{Yoo06-SUS} is an
asymptotically optimal low-complexity scheduler in terms of sum rate
for large number of users in conjunction with ZFBF, we base our
proposal on the same main principle, i.e., the users with larger
channel powers will be selected if the angles between their channel
directions exceed a predetermined threshold. In single-cell systems,
full CSI of all candidate users is required for SUS. In CoMP
systems, we show that the channel orthogonality among users largely
depends on the average channel gains of the users. Based on this
observation, we derive new metrics for scheduling merely using
large-scale fading gains, with which a low-overhead scheduler is
proposed, named large-scale fading gain based user scheduler
(LargeUS). This scheduler can be applied for a two-stage transmit
strategy, where in the first stage, scheduling is performed based
only on the large-scale fading gains of all candidate users and in
the second stage, beamforming is done using  full CSI of the
selected users. Despite that channel asymmetry is exploited to
derive the LargeUS, simulation results show that it performs close
to SUS using full CSI even when the users are located in cell-edge
regions.

In~\cite{Zhang09}, a channel norm-based scheduler was proposed,
where the users with the largest instantaneous channel norms are
selected, which however differs from our method since the proposed
scheduler takes both the orthogonality and receive power into
account.

\emph{Notations:} Boldface upper and lower case letters denote
matrices and row vectors, and standard lower case letters denote
scalars. Superscripts $(\cdot)^T$, $(\cdot)^H$ and $(\cdot)^\dagger$
denote the transpose, the conjugate transpose and the Moore-Penrose
inverse, respectively. $\Re\{\cdot\}$ and $\Im\{\cdot\}$ denote the
real and imaginary parts, $\mathbb{E}\{\cdot\}$ denotes the
expectation operator, and $\cup$ denotes the union between two sets.
$\|\mathbf{a}\|$ denotes the Euclidean norm of a vector
$\mathbf{a}$, $|\mathbf{A}|$ denotes the determinant of a matrix
$\mathbf{A}$, and $\mathrm{diag}\{\mathbf{A}_1,\dots,\mathbf{A}_n\}$
represents the block diagonal matrix with diagonal matrices
$\mathbf{A}_1,\dots,\mathbf{A}_n$.
$\mathbf{x}\sim\mathcal{CN}(\mathbf{0},\boldsymbol{\Sigma})$ denotes
a circularly symmetric complex Gaussian vector $\mathbf{x}$ with
covariance matrix $\boldsymbol{\Sigma}$ and zero mean.
$\mathbb{C}^{m\times n}$ denotes the set of all $m\times n$ complex
matrices. Finally, $\mathbf{I}$ denotes the identity matrix, and
$\mathbf{0}$ denotes the vector of zeros.

Acronyms used throughout the paper are listed in Table I.

\begin{table}\centering
  \renewcommand{\arraystretch}{1.05}
    \begin{threeparttable}[b]
    \caption{Acronyms} \label{T:Acronyms}
    \begin{tabular}{l|l}
        \toprule
        BS  & Base Station \\ \hline
        CDF  & Cumulative Distribution Function \\ \hline
        CoMP\tnote{$\ast$} & Coordinated Multi-point \\ \hline
        CSI  & Channel State Information \\ \hline
        FDD  & Frequency Division Duplex \\ \hline
        JP   & Joint Processing \\ \hline
        LargeUS & Large-scale fading gain based User Scheduler \\ \hline
        LTE  & Long Term Evolution \\ \hline
        MU-MIMO & Multi-user Multi-input Multi-output \\ \hline
        NMSE & Normalized Mean Square Errors \\ \hline
        OFDM & Orthogonal Frequency Division Multiplexing \\ \hline
        PBPC & Per-BS Power Constraints \\ \hline
        PDF  & Probability Density Function \\ \hline
        PF   & Proportional Fair \\ \hline
        RR   & Round Robin \\ \hline
        SFUS & Selective Feedback based User Scheduler \\ \hline
        SINR & Signal-to-Interference plus Noise Ratio\\ \hline
        SNR  & Signal-to-Noise Ratio \\ \hline
        SUS  & Semi-orthogonal User Scheduler \\ \hline
        TDD  & Time Division Duplex \\ \hline
        ZFBF & Zero-forcing Beamforming \\
        \bottomrule
    \end{tabular}
    \begin{tablenotes}
        \baselineskip=0.8\normalbaselineskip
        \item [$\ast$] {\footnotesize{CoMP is also used to mean CoMP-JP for notational simplicity from Section II.}}
    \end{tablenotes}
    \end{threeparttable}
\end{table}

\section{System Model} \label{P:Systemmodel}
Consider one cluster of a downlink CoMP-JP system  consisting of $M$
coordinated cells, each including one BS and $K$ users. Each BS is
equipped with $N_t$ antennas and each user has a single antenna. For
simplicity, we refer to CoMP-JP as CoMP in the rest of the paper.

We consider that all the $MK$ users are located in a cell-edge
region. For each user, we define its local BS as the BS providing
the largest receive power. The channel between a user and its local
BS is called local channel, while the channels between the user and
its cooperating BSs are called cross channels. We use $i_{km}$ to
denote the index of the $k$th user located in cell $m$. Then for
user $i_{km}$ (denoted by $\ms_{i_{km}}$), its channel asymmetry is
reflected by a parameter defined as $\rho_{i_{km}} =
\frac{\alpha_{i_{km}m}}{\max_{n\neq m}\alpha_{i_{km}n}}$, where
$\alpha_{i_{km}m}$ and $\alpha_{i_{km}n}$ for $n\neq m$ are the
large-scale fading gains of local and cross channels of
$\ms_{i_{km}}$, respectively. The large-scale fading gains include
both pathloss and shadowing. Shadowing usually follows a log-normal
distribution, but we assume that $\alpha_{i_{km}n}$ is deterministic
as in most works in the literature of CoMP~\cite{Heath10ICASSP}
because the time scale of user scheduling is much shorter than the
large-scale fading variations. If the value of $\rho_{i_{km}}$ is
less than a predefined parameter $\bar\rho$, we say that
$\ms_{i_{km}}$ is located in a ``$\bar\rho$ cell-edge region''.
Fig.~1 illustrates the 3~dB and 10~dB cell-edge regions.

Let $\tilde{\mathbf{h}}_{i_{km}n}$ and $\mathbf{h}_{i_{km}n}$ denote
the small-scale fading channel and the composite channel from the
$n$th BS (denoted by $\bs_n$) to $\ms_{i_{km}}$, where
$\mathbf{h}_{i_{km}n} =
\sqrt{\alpha_{i_{km}n}}\tilde{\mathbf{h}}_{i_{km}n} \in
\mathbb{C}^{1 \times N_t}$. Then the global channel of
$\ms_{i_{km}}$ can be expressed as $\mathbf{h}_{i_{km}} =
\left[\mathbf{h}_{i_{km}1}, \dots, \mathbf{h}_{i_{km}M} \right] \in
\mathbb{C}^{1 \times MN_t}$. To focus on the impact of channel
asymmetry on user scheduling, we consider flat fading channel and
the small-scale fading channel follows independent and identically
distributed (i.i.d.) complex Gaussian distribution, i.e.,
$\tilde{\mathbf{h}}_{i_{km}n}\sim
\mathcal{CN}(\mathbf{0},\mathbf{I})$. Furthermore, we assume that
the small-scale fading channels from different BSs to each user are
independent. Then $\mathbf{h}_{i_{km}} \sim
\mathcal{CN}(\mathbf{0},{\mathbf{R}}_{i_{km}})$ with
${\mathbf{R}}_{i_{km}} =
\text{diag}\{\alpha_{i_{km}1}{\mathbf{I}},\dots,\alpha_{i_{km}M}
{\mathbf{I}}\}$.

We consider time division duplex (TDD) CoMP systems with a two-stage
transmission strategy. In the first stage, scheduling is performed
using only large-scale fading gains of all the candidate users. In
the second stage, the selected users are informed to provide their
full CSI, with which the precoders are computed.  We employ
Moore-Penrose inverse based ZFBF for precoding, which is of low
complexity and is widely studied~\cite{Yoo06-SUS,Wiesel2008}.

\begin{figure}
\centering
\begin{minipage}[t]{0.48\textwidth}
        \includegraphics[width=1\textwidth]{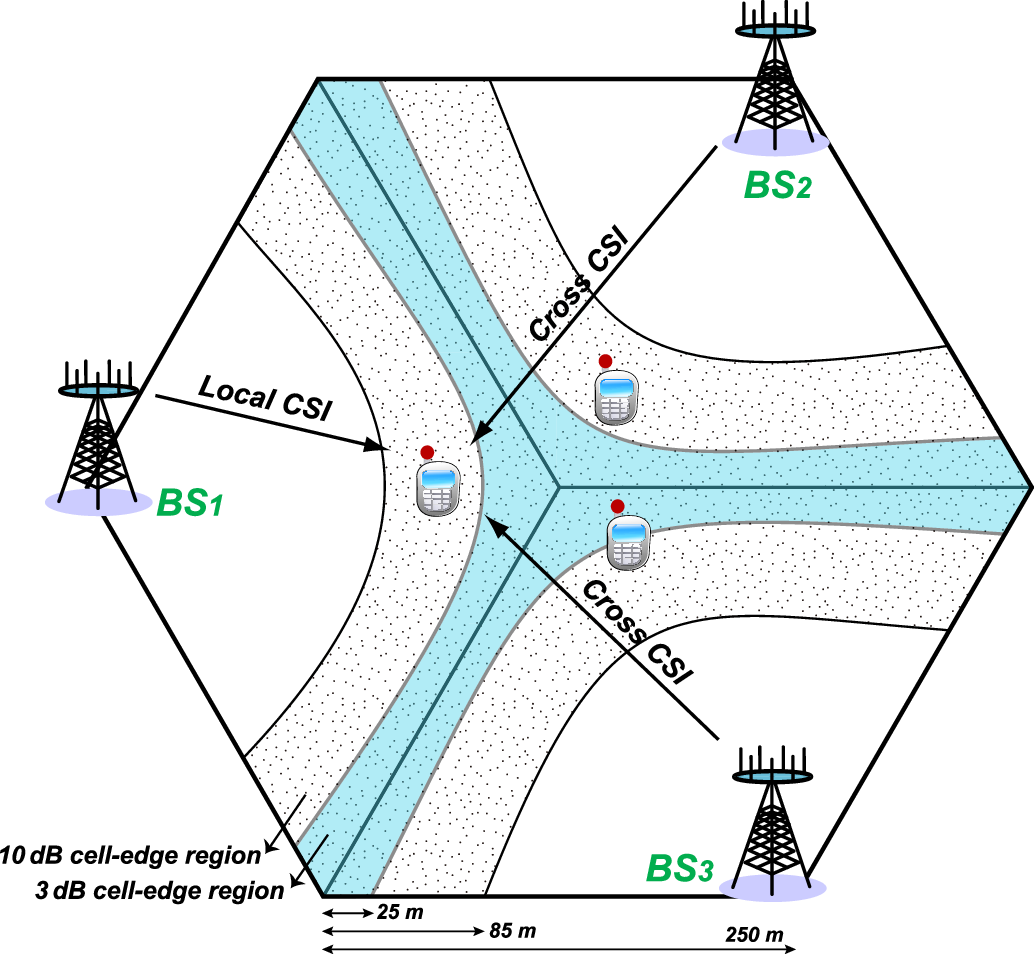}
\end{minipage}
\caption{\label{F:systemmodel} The layout of the reference CoMP
cluster, where three BSs cooperatively serve users located in
cell-edge regions, the cell radius $r$ is 250~m, and the pathloss
follows $35.3+ 37.6\log_{10}(d)$~\cite{TR36.814}. The 3~dB and 10~dB
cell-edge regions are around 25~m and 85~m away from the cell
boundary, respectively. The regular cell-edge regions are obtained
without considering shadowing in order to provide an easy
understanding.}
\end{figure}

\section{Low-overhead User Scheduler}
In this section, we propose a low-overhead scheduler using the same
principle as SUS by exploiting the channel asymmetry. We first show
that the channel orthogonality between users can be approximately
determined based on the average channel gains. Then we derive new
scheduling metrics only with large-scale fading gains, from which a
low-overhead scheduler is developed for CoMP systems.

\subsection{Probability of Orthogonality between CoMP Channels}
The scheduling principle employed in SUS is as follows: the users
with large receive power and with good orthogonality among each
other are successively selected~\cite{Yoo06-SUS}, which is called
SUS principle for short in the sequel. For two users (say $\ms_1$
and $\ms_2$), the orthogonality between their channels can be
represented by the cosine of the angle between their channel vectors
$\cos\theta =
\frac{|\mathbf{h}_{1}\mathbf{h}_{2}^H|}{\|\mathbf{h}_{1}\|\|\mathbf{h}_{2}\|}$,
$0\leq\theta\leq{\pi}/{2}$. When $\cos\theta=0$, the channels of the
two users are~orthogonal.

We derive the probability density function (PDF) of $\cos^2\theta$
in Appendix \ref{A:Jointpdf}. To gain some insight, the PDF for the
special case of two single-antenna coordinated BSs is given by
\begin{align}
\label{E:pdf_cos_2}
    f_{\cos^2\theta}(x)
    =\int_0^1  &
    \frac{\kappa_{1}\kappa_{2}}{\left((\kappa_{1}-1)t_1+1\right)^2\left((1-\kappa_{2})\delta+\kappa_{2}\right)^2}\nonumber\\
    &\cdot\left(1-\frac{4(\kappa_{1}-1)^2t_2}{\left((\kappa_{1}-1)t_1+1\right)^2}\right)^{-\frac{3}{2}}d\delta,
\end{align}
where $t_1=\delta(1-x)+(1-\delta)x$, $t_2=\delta(1-\delta)x(1-x)$,
and $\kappa_i = \alpha_{i1}/\alpha_{i2}$ for $i=1, 2$. When
$\kappa_i = 1$ (i.e., $0$~dB), $\ms_i$ is located at the cell
boundary and its channel is not asymmetric.

According to the results shown in~\cite{Yoo06-SUS}, the cosine of
the angle between channel vectors of any two selected users by SUS
is less than 0.5. Therefore, we can say that $\ms_1$ and $\ms_2$ are
semi-orthogonal when $\cos^2\theta < 0.25$. Numerical results of the
probability of $\cos^2\theta < 0.25$ (i.e., $\theta > 60^\circ$) are
depicted in Fig.~\ref{F:tightness}(a), where two users are
considered. It is shown that when the channel of a user is not
asymmetric, i.e., $\kappa_1 = 0$~dB, the probability that it is
orthogonal with the other user's channel is low and does not depend
on the location of the other user. This implies that in single-cell
systems with i.i.d. channels, we have to use full CSI to decide the
channel orthogonality. When the channels of the two users are
asymmetric, their orthogonality depends on the users' location. The
probability of $\theta
> 60^\circ$ is very low when two users are located in the same cell,
while it becomes larger when they are in different cells. Resembling
single-cell MIMO systems, when the BS has more antennas, the
behavior is similar except that the users will be orthogonal with a
higher probability. This indicates that we can use large-scale
fading gains to decide the orthogonality between users in CoMP
systems. In the following, we derive low-overhead scheduling metrics
for selecting users based on the SUS principle by efficiently
exploiting the large-scale fading gains.

\subsection{Low-overhead SUS-based Scheduler}
\subsubsection{{Scheduling Metrics with Large-scale Fading Gains}} When full CSI of all candidate users is available, to decide
whether $\ms_{i_{km}}$ should be selected in the $(l+1)$th iteration
with the SUS principle~\cite{Yoo06-SUS}, we need to compute the
orthogonal projection power of its channel vector on a subspace
spanned by the channel vectors of already selected users, i.e.,
\begin{equation} \label{E:Normproj}
    \nu_{\mathcal{S}_li_{km}} = \mathbf{h}_{i_{km}}\mathbf{Q}_{\mathcal{S}_{l}}^\bot\mathbf{h}_{i_{km}}^H,
\end{equation}
where $\mathbf{Q}_{\mathcal{S}_{l}}^\bot
=\mathbf{I}-\mathbf{H}_{\mathcal{S}_{l}}^H
                (\mathbf{H}_{\mathcal{S}_{l}}\mathbf{H}_{\mathcal{S}_{l}}^H)^{\dagger}\mathbf{H}_{\mathcal{S}_{l}}$ is the orthogonal
projection matrix onto the subspace spanned by
$\mathbf{H}_{\mathcal{S}_{l}}$, $\mathcal{S}_{l} =
\{s_1,\dots,s_l\}$ is the scheduling result before the $(l+1)$th
iteration, and $\mathbf{H}_{\mathcal{S}_{l}} = [\mathbf{h}_{s_1}^T,
\dots, \mathbf{h}_{s_l}^T]^T\in\mathbb{C}^{l\times MN_t}$ with
$\mathbf{h}_{s_i} = [\mathbf{h}_{s_i1}, \dots,
\mathbf{h}_{s_iM}]\in\mathbb{C}^{1\times MN_t}$ for $i = 1,\dots,l$.
We also need to obtain the metric reflecting the orthogonality
between $\ms_{i_{km}}$ and already selected users in
$\mathcal{S}_{l}$, which is defined as
\begin{equation} \label{E:angleSUS}
\mu_{\mathcal{S}_li_{km}}
=\frac{\nu_{\mathcal{S}_li_{km}}}{\|\mathbf{h}_{i_{km}}\|^2}.
\end{equation}
According to the SUS principle, $\ms_{i_{km}}$ will be selected if
$\mu_{\mathcal{S}_li_{km}}$ is larger than a specific orthogonal
threshold and $\nu_{\mathcal{S}_li_{km}}$ is larger than those of
other candidate users.

\begin{figure}
\centering
\begin{minipage}[t]{0.48\textwidth}
        \includegraphics[width=1\textwidth]{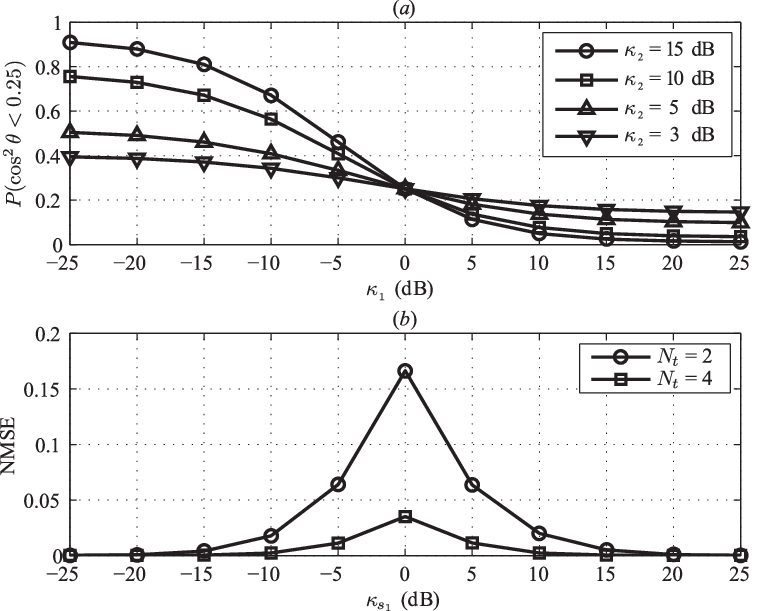}
\end{minipage}
\caption{\label{F:tightness} $(a)$ The probability of
$\cos^2\theta<0.25$ (i.e., $\theta>60^\circ$) as a function of
$\kappa_{1}$ in dB with $N_t=1$. $\ms_1$ and $\ms_2$ are two
arbitrary users located in two cooperative cells. $\ms_{i}$ is
located in cell 1 if $\kappa_i > 0$~dB and in cell 2 if
$\kappa_i<0$~dB. $(b)$ The NMSE between $\nu_{\mathcal{S}_li_{km}}$
and its approximation $\nu_{\mathcal{S}_li_{km}}^{app}$ as a
function of $\kappa_{s_1}$ with $N_t=2, 4$, where $\mathcal{S}_l$
includes only one already selected user $\ms_{s_1}$ and
$\ms_{i_{km}}$ is an arbitrary candidate user uniformly distributed
in two cells. The results are averaged over 10000 small-scale
Rayleigh fading channels following complex Gaussian distribution
$\mathcal{CN}(\mathbf{0}, \mathbf{I})$.}
\end{figure}

When only large-scale fading gains are available, the scheduling
metrics $\mu_{\mathcal{S}_li_{km}}$ and $\nu_{\mathcal{S}_li_{km}}$
cannot be used any more.

To derive the metrics only dependent on average channel gains, we
take the expectation of $\nu_{\mathcal{S}_li_{km}}$ over the unknown
small-scale channels. However, the expectation of
$\nu_{\mathcal{S}_li_{km}}$ is very hard to derive, because the
global channel is non-identically distributed due to the channel
asymmetry even when the channel from each BS is i.i.d.. To
circumvent this problem, we first provide an approximation of
$\nu_{\mathcal{S}_li_{km}}$ by exploiting the channel asymmetry. In
particular, considering that the cross channels have smaller average
gains than local channels, we approximate the cross channels of the
already selected users in $\mathcal{S}_l$ as zeros. Then the matrix
$\mathbf{H}_{\mathcal{S}_{l}}$ can be approximated as a sparse
matrix, denoted by
$\mathbf{H}_{\mathcal{S}_{l}}^{app}\in\mathbb{C}^{l\times MN_t}$. By
performing row interchange operations on the sparse matrix, we can
obtain a block-diagonal matrix, i.e.,
\begin{equation} \label{E:app-diag}
  \mathbf{\hat H}_{\mathcal{S}_{l}} =
\boldsymbol{\Gamma}_{\mathcal{S}_{l}}\mathbf{H}^{app}_{\mathcal{S}_{l}}
\triangleq\text{diag}\{\mathbf{H}_{\mathcal{S}_{l},1},
\dots,\mathbf{H}_{\mathcal{S}_{l},M}\},
\end{equation}
where $\boldsymbol{\Gamma}_{\mathcal{S}_{l}}\in\mathbb{C}^{l\times
l}$ is the row permutation  matrix satisfying
$\boldsymbol{\Gamma}_{\mathcal{S}_{l}}^H\boldsymbol{\Gamma}_{\mathcal{S}_{l}}
= \mathbf{I}$. Let $I_{\mathcal{S}_l,n}$ denote the number of
already selected users that are located in cell $n$. Then the
diagonal block of $\mathbf{\hat H}_{\mathcal{S}_{l}}$,
$\mathbf{H}_{\mathcal{S}_{l},n}$, has the dimension of
$I_{\mathcal{S}_l,n}\times N_t$, and consists of local channels of
the $I_{\mathcal{S}_l,n}$ selected users located in cell $n$ for $n
= 1,\dots, M$.

{\emph{\textbf{Example:}}} \emph{Consider that two cells are
coordinated and three users in $\mathcal{S}_l = \{s_1,s_2,s_3\}$ are
successively scheduled, i.e., $M=2$ and $l=3$. Suppose that the
first user $\ms_{s_1}$ and the third user $\ms_{s_3}$ are located in
cell 2, and the second user $\ms_{s_2}$ is located in cell 1. Then,
$I_{\mathcal{S}_l,1} = 1$ and $I_{\mathcal{S}_l,2} = 2$. Following
the definition of $\mathbf{H}_{\mathcal{S}_l}$ after
\eqref{E:Normproj}, we have
\begin{equation*}
  \mathbf{H}_{\mathcal{S}_l} = \left[
    \begin{mmatrix}
        \mathbf{h}_{s_11} & \underline{\mathbf{h}_{s_12}}\\
        \underline{\mathbf{h}_{s_21}} & \mathbf{h}_{s_22}\\
        \mathbf{h}_{s_31} & \underline{\mathbf{h}_{s_32}}
    \end{mmatrix}
  \right]\in\mathbb{C}^{3\times 2N_t},
\end{equation*}
where the local channel of each scheduled user is underlined. Then
the approximation of $\mathbf{H}_{\mathcal{S}_l}$,
$\mathbf{H}_{\mathcal{S}_l}^{app}$, can be obtained by omitting the
cross channels of the scheduled users as
\begin{equation*} \label{E:1}
  \mathbf{H}_{\mathcal{S}_l}\approx\mathbf{H}_{\mathcal{S}_l}^{app} = \left[
    \begin{mmatrix}
        \mathbf{0} & \underline{\mathbf{h}_{s_12}}\\
        \underline{\mathbf{h}_{s_21}} & \mathbf{0}\\
        \mathbf{0} & \underline{\mathbf{h}_{s_32}}
    \end{mmatrix}
  \right]\in\mathbb{C}^{3\times 2N_t}.  
\end{equation*}
Further by performing row interchange operations on
$\mathbf{H}_{\mathcal{S}_l}^{app}$, we obtain the block-diagonal
matrix
\begin{equation*}\label{E:2}
  \mathbf{\hat H}_{\mathcal{S}_l} = \boldsymbol{\Gamma}_{\mathcal{S}_{l}}\mathbf{H}_{\mathcal{S}_l}^{app} =  \left[
    \begin{mmatrix}
        \underline{\mathbf{h}_{s_21}} & \mathbf{0}\\
        \mathbf{0} & \underline{\mathbf{h}_{s_12}}\\
        \mathbf{0} & \underline{\mathbf{h}_{s_32}}
    \end{mmatrix}
  \right]\in\mathbb{C}^{3\times 2N_t}, 
\end{equation*}
where  $\boldsymbol{\Gamma}_{\mathcal{S}_{l}}= \left[
\begin{smallmatrix}
        0 & 1 & 0\\
        1 & 0 & 0\\
        0 & 0 & 1
    \end{smallmatrix}
  \right]$ is a row permutation matrix, and the two diagonal blocks $\mathbf{H}_{\mathcal{S}_l,1} =
\mathbf{h}_{s_21}$ and $\mathbf{H}_{\mathcal{S}_l,2} =
\left[\begin{smallmatrix} \mathbf{h}_{s_12}\\
\mathbf{h}_{s_32}
\end{smallmatrix}\right]$, which have the dimension of $1\times N_t$ and $2\times
N_t$, respectively.  \quad \ $\Box$}

Since $\mathbf{H}_{\mathcal{S}_{l}} \approx
\mathbf{H}_{\mathcal{S}_{l}}^{app}$, we obtain from
(\ref{E:app-diag}) that $\mathbf{H}_{\mathcal{S}_{l}} \approx
\boldsymbol{\Gamma}_{\mathcal{S}_{l}}^H\mathbf{\hat
H}_{\mathcal{S}_{l}} $. Therefore $\nu_{\mathcal{S}_li_{km}}$ in
(\ref{E:Normproj}) can be approximated as
\begin{align} \label{E:Normproj-blockdiagonal}
&\nu_{\mathcal{S}_li_{km}} \approx
\mathbf{h}_{i_{km}}\left(\mathbf{I} -\mathbf{\hat
H}_{\mathcal{S}_{l}}^H\boldsymbol{\Gamma}_{\mathcal{S}_{l}}
(\boldsymbol{\Gamma}_{\mathcal{S}_{l}}^H\mathbf{\hat
H}_{\mathcal{S}_{l}}\mathbf{\hat
H}_{\mathcal{S}_{l}}^H\boldsymbol{\Gamma}_{\mathcal{S}_{l}})^{\dagger}
\boldsymbol{\Gamma}_{\mathcal{S}_{l}}^H\mathbf{\hat H}_{\mathcal{S}_{l}}\right)\mathbf{h}_{i_{km}}^H \nonumber\\
&= \|\mathbf{h}_{i_{km}}\|^2 - \sum_{n=1}^M \mathbf{h}_{i_{km}n}
\mathbf{H}_{\mathcal{S}_{l},n}^H\left(\mathbf{H}_{\mathcal{S}_{l},n}
    \mathbf{H}_{\mathcal{S}_{l},n}^H\right)^{\dagger}\mathbf{H}_{\mathcal{S}_{l},n}\mathbf{h}_{i_{km}n}^H\nonumber\\
    &\triangleq \nu_{\mathcal{S}_li_{km}}^{app}.
\end{align}

The accuracy of the approximation is shown in
Fig.~\ref{F:tightness}(b). Considering the case of $M=2$ and $l=1$,
we evaluate the normalized mean square errors (NMSE) between
$\nu_{\mathcal{S}_li_{km}}$ and $\nu_{\mathcal{S}_li_{km}}^{app}$,
defined as
$\mathbb{E}\{|\nu_{\mathcal{S}_li_{km}}^{app}-\nu_{\mathcal{S}_li_{km}}|^2\}/\mathbb{E}\{|\nu_{\mathcal{S}_li_{km}}|^2\}$,
as a function of the location of the already selected user
$\ms_{s_1}$, where $\ms_{i_{km}}$ is an arbitrary candidate user
uniformly distributed in the two cells. It can be seen from the
simulation results that the largest NMSE occurs when $\kappa_{s_1} =
0$~dB, i.e., the global channel of $\ms_{s_1}$ is symmetric so that
it can be regarded as a single-cell channel with $MN_t$ antennas.
The NMSE rapidly decreases when the CoMP channel of the selected
user becomes asymmetric. For instance, compared with the largest
NMSE at $\kappa_{s_1} = 0$~dB when $N_t=2$, over 60\% reduction of
NMSE can be observed at $\kappa_{s_1} = \pm 5$~dB. This indicates
that the approximation is still sufficiently accurate when the users
are located in cell-edge regions, where the difference between the
local and cross channel gains is not large. Moreover, the accuracy
of the approximation improves with the increase of antenna number
$N_t$. This can be explained as follows. To obtain
$\nu_{\mathcal{S}_li_{km}}^{app}$, the approximation is made only
for the channel matrix of already selected users
$\mathbf{H}_{\mathcal{S}_{l}}$ but not for the channel of candidate
user $\mathbf{h}_{i_{km}}$. According to the SUS principle, the
first selected users should have large receive power and be more
orthogonal among each other. This implies that these users are
relatively close to their local BSs. As a result, their cross
channels can be approximated as zeros and the subspace spanned by
$\mathbf{H}_{\mathcal{S}_{l}}$ can be approximated by that spanned
by $\hat{\mathbf{H}}_{\mathcal{S}_{l}}$. Since when each BS has more
antennas the channels will on average become more orthogonal, the
approximation is more accurate when $N_t$ increases.

We next derive the expectation of $\nu_{\mathcal{S}_li_{km}}^{app}$
over unknown small-scale channels. In
(\ref{E:Normproj-blockdiagonal}), the term
$\mathbf{H}_{\mathcal{S}_{l},n}^H\left(\mathbf{H}_{\mathcal{S}_{l},n}
\mathbf{H}_{\mathcal{S}_{l},n}^H\right)^{\dagger}\mathbf{H}_{\mathcal{S}_{l},n}$
is a projection matrix. Therefore, considering $\mathbf{h}_{i_{km}n}
\sim \mathcal{CN}(\mathbf{0}, \alpha_{i_{km}n}\mathbf{I})$, it is
readily obtained that
\begin{align*}
&\mathbb{E}\left\{\mathbf{h}_{i_{km}n}
\mathbf{H}_{\mathcal{S}_{l},n}^H\left(\mathbf{H}_{\mathcal{S}_{l},n}
\mathbf{H}_{\mathcal{S}_{l},n}^H\right)^{\dagger}\mathbf{H}_{\mathcal{S}_{l},n}\mathbf{h}_{i_{km}n}^H\right\}\nonumber\\
&=
\begin{cases}
I_{\mathcal{S}_l,n}\alpha_{i_{km}n} & \text{if $I_{\mathcal{S}_l,n} < N_t$,}\\
N_t\alpha_{i_{km}n} & \text{if $I_{\mathcal{S}_l,n} \geq N_t$.}
\end{cases}
\end{align*}
Further considering that $\mathbb{E}\{\|\mathbf{h}_{i_{km}}\|^2\} =
N_t\sum_{n=1}^M\alpha_{i_{km}n}$, the expectation of
$\nu_{\mathcal{S}_li_{km}}^{app}$ can be obtained~as
\begin{equation} \label{E:Average-Normproj-blockdiagonal}
 \mathbb{E}\{\nu_{\mathcal{S}_li_{km}}^{app}\} = \sum_{n=1}^M (N_t
- I_{\mathcal{S}_l,n})^{+}\alpha_{i_{km}n} \triangleq
    \bar\nu_{\mathcal{S}_li_{km}}^{app},
\end{equation}
where $(x)^{+} = \max(x,0)$.

The metric to reflect orthogonality only using large-scale fading
gains can also be derived. By replacing $\nu_{\mathcal{S}_li_{km}}$
and $\|\mathbf{h}_{i_{km}}\|^2$ in (\ref{E:angleSUS}) with their
expectations, we obtain the metric~as
\begin{equation} \label{E:Angle-sine}
 \bar\mu_{\mathcal{S}_li_{km}}^{app}
=\frac{\bar\nu_{\mathcal{S}_li_{km}}^{app}}{N_t\sum_{n=1}^M\alpha_{i_{km}n}}.
\end{equation}

The scheduling metrics $\bar\nu_{\mathcal{S}_li_{km}}^{app}$ and
$\bar\mu_{\mathcal{S}_li_{km}}^{app}$ depend on the locations of
already selected users. We take an example to illustrate the
effectiveness of using the metric  of
$\bar\nu_{\mathcal{S}_li_{km}}^{app}$ for scheduling in CoMP
systems. We apply the SUS principle by using
$\bar\nu_{\mathcal{S}_li_{km}}^{app}$ in a scenario of two-cell
CoMP. Let $\ms_0$ be the first selected user, and $\ms_1$ and
$\ms_2$ be the two candidate users in the second iteration
satisfying semi-orthogonal constraints. Assume that $\ms_1$ and
$\ms_2$ are symmetrically located in cell 1 and cell 2, i.e., they
have the same average local channel gains $\alpha_{11} = \alpha_{22}
\triangleq \alpha_L$ and cross channel gains $\alpha_{12} =
\alpha_{21} \triangleq \alpha_C$, where $\alpha_L>\alpha_C$. Then if
$\ms_0$ is located in cell 2, we can obtain from
(\ref{E:Average-Normproj-blockdiagonal}) that
$\bar\nu_{\mathcal{S}_l1}^{app} = N_t\alpha_L + (N_t-1)\alpha_C$ and
$\bar\nu_{\mathcal{S}_l2}^{app} = (N_t-1)\alpha_L + N_t\alpha_C$. So
$\ms_1$ located in different cell from $\ms_0$ will be the second
selected user since
$\bar\nu_{\mathcal{S}_l1}^{app}>\bar\nu_{\mathcal{S}_l2}^{app}$. The
result is consistent with the observation from
Fig.~\ref{F:tightness}(a) that the users in different cells are more
orthogonal, which should be scheduled simultaneously. Note that
channel asymmetry is essential for the effectiveness of the metric.
Otherwise, if $\alpha _L = \alpha_C$ in the example,
$\bar\nu_{\mathcal{S}_l1}^{app}$ will equal to
$\bar\nu_{\mathcal{S}_l2}^{app}$. Then the scheduler cannot judge
which user should be selected from the metric.

\subsubsection{Low-overhead Scheduler} With the scheduling metrics ${\bar\nu}_{\mathcal{S}_li_{km}}^{app}$ and ${\bar\mu}_{\mathcal{S}_li_{km}}^{app}$,
we propose a low-overhead large-scale fading gain based user
scheduler (LargeUS) based on the SUS principle, which operates as
follows.

\begin{enumerate}
    \item[(1)] Select the user with the maximum average channel gain normalized by noise as the first user, i.e.,

            \begin{equation} \label{E:NUS1}
            s_1 = \arg \underset{i_{km}\in\mathcal{T}_0}{
                \max}~ \frac{{N_t\sum_{n=1}^{ M}\alpha_{i_{km}n}}}{\sigma_{i_{km}}^2},
            \end{equation}
            where $\sigma_{i_{km}}^2$ is the variance of noise and $\mathcal{T}_0=\{1,2,\ldots,MK\}$.
            Let $\mathcal{T}_l$ and $\mathcal{S}_l$ denote the user pool and
            the scheduling result at the $l$th step, and set $\mathcal{S}_1=\{s_1\}$ and $l = 1$.
    \item[(2)] When $l \leq \min(MN_t, MK)$, obtain the user pool as
            \begin{equation} \label{E:T_l}
                \mathcal{T}_l =
                \big\{i_{km} \in \mathcal{T}_{l-1}, i_{km} \notin \mathcal{S}_l \ |\ {\bar\mu}_{\mathcal{S}_li_{km}}^{app} \geq 1- \epsilon
                \big\},
            \end{equation}
            where $\epsilon$ is a specific orthogonality threshold.
            If $\mathcal{T}_l = \phi$ \  (empty set), the iteration will stop. Otherwise, compute
            ${\bar\nu}_{\mathcal{S}_li_{km}}^{app}$ and select the new user
            as
            \begin{equation} \label{E:NUS2}
                s_{l+1} = \arg \underset{i_{km}\in\mathcal{T}_l}{
                \max}~ \frac{{\bar\nu}_{\mathcal{S}_li_{km}}^{app}}{\sigma_{i_{km}}^2}.
            \end{equation}
            Set $\mathcal{S}_{l+1} = \mathcal{S}_{l}\cup\{s_{l+1}\}$ and $l = l+1$. $\qquad \qquad  \qquad \ \ \square$
\end{enumerate}

As mentioned before, LargeUS is applicable for a two-stage
transmission strategy. After scheduling among all candidate users
with their average channel gains in the first stage, the selected
users are informed to provide full CSI for precoding in the second
stage. Since only large-scale fading gains are required for LargeUS,
which can be obtained either by long-term feedback from users or by
averaging over past received signals at the BSs, uplink training is
only necessary for precoding during the second stage.\footnote{The
LargeUS successively selects users following the same procedure as
SUS~\cite{Yoo06-SUS}, but SUS requires full knowledge of the global
channels of all candidate users. SUS is performed by replacing
$N_t\alpha_{i_{km}n}$ in (\ref{E:NUS1}) with
$\|\mathbf{h}_{i_{km}n}\|^2$, replacing
${\bar\mu}_{\mathcal{S}_li_{km}}^{app}$ in (\ref{E:T_l}) with
${\mu}_{\mathcal{S}_li_{km}}$ defined in (\ref{E:angleSUS}), and
replacing ${\bar\nu}_{\mathcal{S}_li_{km}}^{app}$ in (\ref{E:NUS2})
with ${\nu}_{\mathcal{S}_li_{km}}$ defined in (\ref{E:Normproj}).
After scheduling, the ZFBF precoder can be used for both LargeUS and
SUS when full CSI of the selected users is available.}

In the precoding stage, the Moore-Penrose inverse based
ZFBF~\cite{Yoo06-SUS} is employed for downlink transmission to the
co-scheduled users in the same time and frequency resources. Let
$\mathcal{S}_L = \{s_1, \dots, s_L\}$ denote the $L$ finally
scheduled users. For $\ms_{s_l}$, its precoding vector is comprised
of a unit norm beamforming vector $\mathbf{w}_{s_l}$ and the
transmit power $p_{s_l}$. The beamforming vector can be expressed as
\begin{equation} \label{E:GeneralZF}
  \mathbf{w}_{s_l} =
  \frac{\mathbf{h}_{s_l}\mathbf{Q}_{\mathcal{S}_{\bar l}}^\bot}{\|\mathbf{h}_{s_l}\mathbf{Q}_{\mathcal{S}_{\bar l}}^\bot\|},
\end{equation}
where $\mathcal{S}_{\bar l} = \{j\in\mathcal{S}_L, j\neq s_l\}$
includes all selected users except for $\ms_{s_l}$, and
$\mathbf{Q}_{\mathcal{S}_{\bar l}}^\bot$ is the orthogonal
projection matrix onto the subspace spanned by the channels of users
in $\mathcal{S}_{\bar l}$. Since power cannot be shared among
coordinated BSs, the per-BS power constraint (PBPC) should be
satisfied for power allocation. Given the beamforming vectors
$\mathbf{w}_{s_l}$, the optimal power allocation $p_{s_l}$, aimed at
maximizing sum rate, can be numerically obtained by using the method
in~\cite{Wiesel2008}.

\subsubsection{Threshold Selection} We next discuss the selection of
orthogonality threshold $\epsilon$ by analyzing its impact on
performance. To connect the threshold selection with the
performance, the effective channel gain of $\ms_{s_l}$ normalized by
noise was considered in~\cite{Yoo06-SUS}, which is
\begin{equation} \label{E:GeneralZF1}
  \gamma_{s_{l}} = \frac{|\mathbf{h}_{s_l}\mathbf{w}_{s_l}^H|^2}{\sigma_{s_{l}}^2}=
  \frac{\mathbf{h}_{s_l}\mathbf{Q}_{\mathcal{S}_{\bar l}}^\bot\mathbf{h}_{s_l}^H}{\sigma_{s_{l}}^2}.
\end{equation}
Since we use average channel gains for scheduling, we need to employ
the average normalized effective channel gain, which can be obtained
from (\ref{E:Normproj}), (\ref{E:Normproj-blockdiagonal}) and
(\ref{E:Average-Normproj-blockdiagonal}) as
\begin{equation} \label{E:NUS-SNRUB-1}
  \mathbb{E}\{\gamma_{s_{l}}\} \approx \frac{\sum_{n=1}^M (N_t -
    I_{\mathcal{S}_{\bar l},n})^{+}\alpha_{s_{l}n}}{\sigma_{s_{l}}^2}.
\end{equation}

The orthogonality threshold $\epsilon$ has an intertwined impact on
the performance. On one hand, $\mathbb{E}\{\gamma_{s_{l}}\}$
increases with the decrease of the number of selected users
$I_{\mathcal{S}_{\bar l},n}$. Therefore, a small threshold
$\epsilon$ is preferred to reduce the size of the user pool as shown
in (\ref{E:T_l}). On the other hand, multiuser diversity gain
achieved by exploiting the large-scale channel difference among
users depends on the size of $\mathcal{T}_{l-1}$, from which
$\ms_{s_{l}}$ is selected.
Hence a sufficiently large threshold should be chosen to ensure a
large user pool. We will evaluate the impact of $\epsilon$ via
simulations in Section IV.

\subsubsection{Training Overhead} \label{S:Trainingoverhead} Compared
with SUS that requires full CSI, the proposed scheduler needs only
large-scale fading gains.

For a TDD CoMP system, by exploiting uplink-downlink channel
reciprocity, the BSs can obtain the total $MN_t$ downlink channel
coefficients from $M$ BSs to one user when the user broadcasts a
single uplink training signal. Therefore, in order to obtain the
full global channels of all $MK$ users for SUS, the orthogonal (in
frequency, time or code domain) training sequences employed by the
system require $MK$-dimensional resources. For LargeUS, the users
are scheduled only using large-scale fading gains, which can be
estimated by each user and fed back to its local BS. The overhead
for conveying the long-term information is negligible.
Alternatively, they can also be estimated at the BSs by averaging
over the received signals. The uplink training is only used for
estimating the full global channels of $L$ scheduled users for
precoding in the second stage, which requires $L$-dimensional
resources.

It should be pointed out that at the system level, the reduction of
the overall uplink training overhead achieved by the proposed
scheduler depends on the number of users participating CoMP
transmission. The overall gain will be significant if the cells are
densely deployed such that most users prefer to be served with CoMP.

\section{Simulation Results}
We evaluate the performance of the proposed scheduler via
simulations. We consider a homogeneous cellular network and focus on
the performance of a reference cooperative cluster consisting of
three coordinated cells. We model the interference from surrounding
non-cooperative cells as white noise, which is the worst-case
interference and results in pessimistic performance~\cite{Huang09}.
The layout of the reference cooperative cluster is shown in
Fig.~\ref{F:systemmodel}. In each cell $K=10$ users are uniformly
distributed in a $\bar\rho$ cell-edge region, where $\bar \rho =
3$~dB and 10~dB are considered in simulations. The cell radius $r$
is set to 250~m, and the average receive signal-to-noise ratio (SNR)
of the users located at cell boundary, $\text{SNR}_{0}$, is set to
0~dB and 10~dB.\footnote{The modeled white noise includes both the
interference from non-cooperative cells and thermal noise. If only
consider thermal noise, $\text{SNR}_0$ will be higher than 20~dB
considering the typical system configurations of
LTE~\cite{TR36.814}. When the inter-cell interference is treated as
noise, the SNR is actually a signal-to-interference plus noise ratio
(SINR) and is much lower. In non-CoMP systems, as shown
in~\cite{TR36.814}, $\text{SNR}_0$ can be as low as -5~dB. In CoMP
systems, however, since the strong interference can be eliminated,
$\text{SNR}_0$ will become higher.} The average receive SNR of a
user from a BS with distance $d$ is computed as $\text{SNR}_{0} +
37.6\log_{10}(\frac{r}{d})$. In order to obtain regular cell-edge
regions that are easy to understand, shadowing is not considered in
simulations. Since shadowing enhances channel asymmetry, the
analysis results are valid for practical channels with shadowing.
The spatially correlated small-scale fading channels are considered
based on the ``Spatial Channel Model" (SCM) in urban macro scenario
with four-wavelength antenna spacing and two-degree angle spread at
each BS~\cite{TR25.996}. Although this model produces frequency
selective channels, we consider a frequency-flat version of the
channel corresponding to a single subcarrier in a subband of an
orthogonal frequency division multiplexing (OFDM) system. For the
scenarios when the subband width is less than the coherence
bandwidth of the channel, the performance over a single subcarrier
can stand for that over a subband. The subbands in a CoMP OFDM
system will be orthogonal with the assumption of perfect
time-frequency synchronization among the coordinated BSs. The
scheduling and precoding on different subbands can be separately
conducted, therefore the obtained results can reflect the
performance of general OFDM systems.

Fairness among users is critical for CoMP systems, which can be
ensured by either Round-Robin (RR) scheduling or proportional fair
(PF) scheduling when full CSI is available. With only large-scale
fading gains, however, applying PF scheduling is not straightforward
because it requires the estimation of user data rate. In the
simulations, we apply LargeUS in a RR fashion similar
to~\cite{Yoo06-SUS}, named RR-LargeUS. Specifically, it selects a
group of users at each time slot based on LargeUS, and removes the
selected users from the user pool at next time slot. This procedure
is repeated until no users are left. As a performance baseline, SUS
with full CSI in a RR fashion (denoted by RR-SUS) is simulated for
both CoMP and non-CoMP systems. For comparison, we also simulate the
selective feedback based user scheduler (SFUS) in a RR fashion
(denoted by RR-SFUS), which will be described in detail later. We
use achievable data rate as performance metric and employ ZFBF for
all schemes. With ZFBF, the achievable data rate of $\ms_{s_l}$ can
be obtained as
$\log(1+\frac{p_{s_l}|\mathbf{h}_{s_l}\mathbf{w}_{s_l}^H|^2}{\sigma_{s_l}^2})$,
where $\mathbf{w}_{s_l}$ is the beamforming vector defined in
(\ref{E:GeneralZF}) and $p_{s_l}$ is the power allocated to
$\ms_{s_l}$ that is optimized aimed at maximizing the achievable sum
rate under PBPC~\cite{Wiesel2008}.

\begin{figure}
\centering
\begin{minipage}[t]{0.48\textwidth}
        \includegraphics[width=1\textwidth]{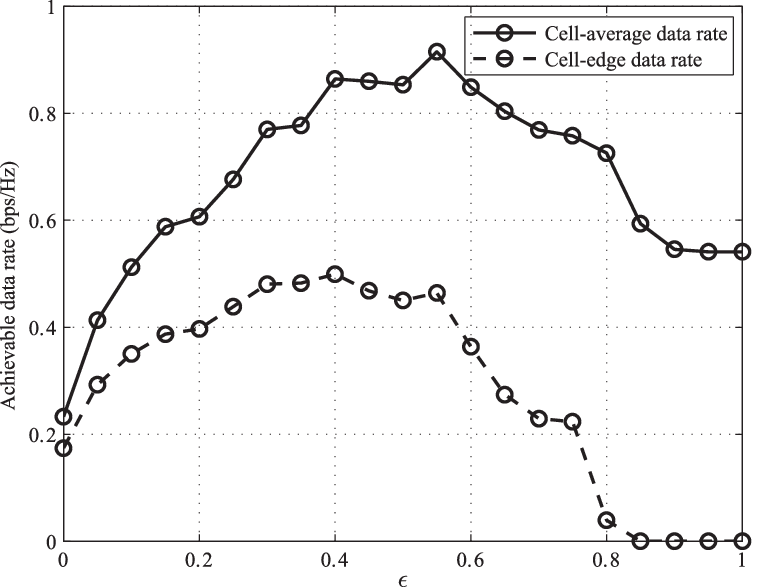}
\end{minipage}
\caption{\label{F:Threshold} Cell-average and cell-edge data rate of
RR-LargeUS as a function of threshold $\epsilon$ with $N_t=4$ and
$\text{SNR}_0 = 10$~dB. The users are located in a 10~dB cell-edge
region.}
\end{figure}

In Fig.~\ref{F:Threshold}, we plot the cell-average data rate and
the cell-edge data rate achieved by the proposed scheduler versus
the threshold $\epsilon$ with $N_t=4$ and $\text{SNR}_0 = 10$~dB.
The cell-average data rate is the average achievable data rate of
all users, and the cell-edge data rate is defined as the 5\% point
of the cumulative distribution function (CDF) of user achievable
data rate. As can be seen, the data rate is not a monotonic function
of the threshold. This agrees with the previous analysis of the
influence of the threshold both on the effective channel gain and on
the multiuser diversity gain. The curves are not smooth because of
the discontinuous changes of the RR scheduling period. Since the
users are served only once during a RR scheduling period, their data
rate is normalized by the period. For a given number of total users,
the scheduling period depends on the number of selected users at
each time slot~\cite{Yoo06-SUS}, which is determined by the
threshold. We can see that the optimal thresholds for maximum
cell-average and cell-edge data rate differ. This is due to the fact
that the orthogonality among users depends on their locations. In
the following simulations we will choose the thresholds that provide
a balance between high cell-average and cell-edge data rate, which
values are given in the figure captions.

\begin{figure*}[!t]
\centering \subfigure[3~dB cell-edge
region]{\includegraphics[width=0.47\textwidth]{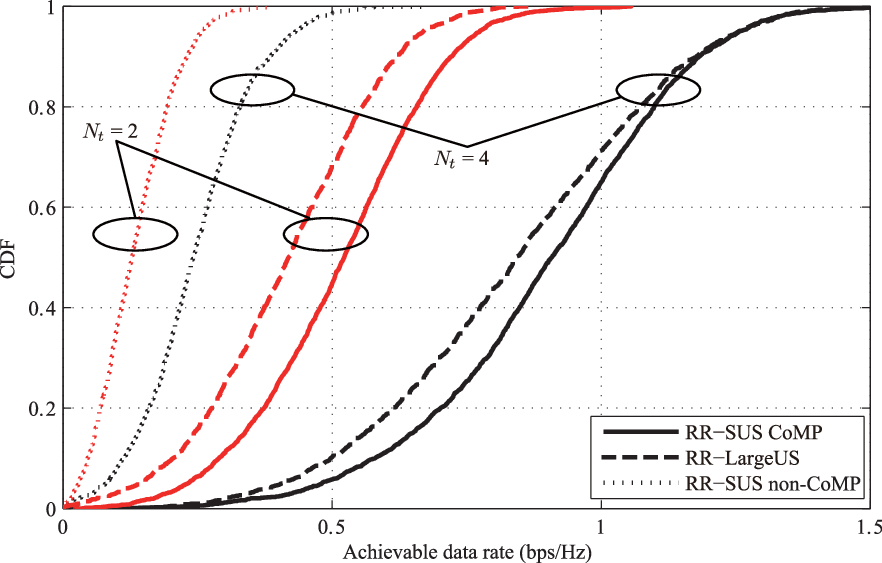}
\label{F:SHO3dB}} \subfigure[10~dB cell-edge
region]{\includegraphics[width=0.47\textwidth]{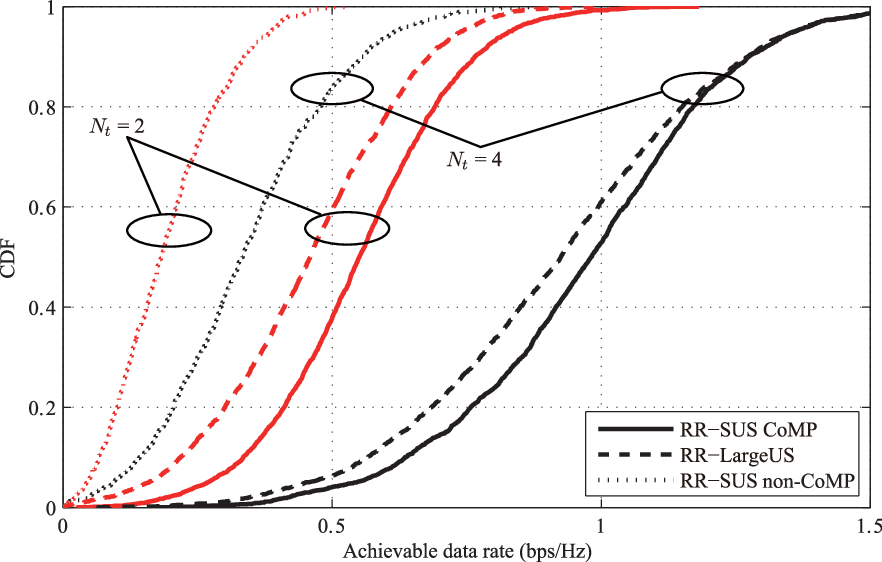}
\label{F:SHO10dB}} \caption{The CDF of user data rate for various
cell-edge regions with $N_t = 2, 4$ and $\text{SNR}_0 = 10$~dB. The
thresholds $\epsilon$ for RR-SUS and RR-LargeUS are chosen as 0.6
and 0.55, respectively, which are also used in Fig. \ref{F:SFUS}.
The legends ``RR-SUS CoMP" and ``RR-SUS non-CoMP" respectively stand
for RR-SUS under CoMP and non-CoMP systems, all with full CSI.}
\label{F:cdfperfect-10dB}
\end{figure*}

\begin{figure*}[!t]
\centering \subfigure[3~dB cell-edge
region]{\includegraphics[width=0.47\textwidth]{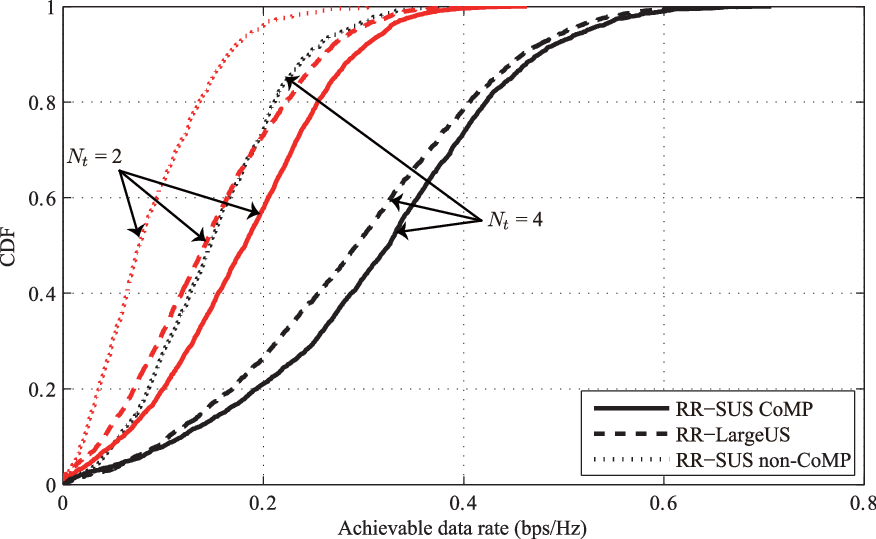}
\label{F:SHO3dB}} \subfigure[10~dB cell-edge
region]{\includegraphics[width=0.47\textwidth]{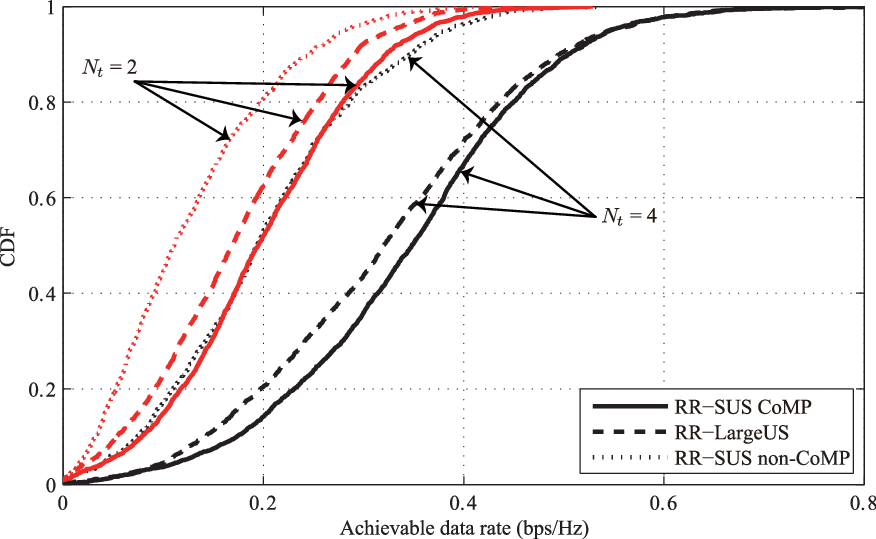}
\label{F:SHO10dB}} \caption{{The CDF of user data rate for various
cell-edge regions with $N_t = 2, 4$ and $\text{SNR}_0 = 0$~dB. The
thresholds $\epsilon$ for RR-SUS and RR-LargeUS are chosen as 0.5
and 0.45, respectively, which are also used in Fig. \ref{F:SFUS-2}.
The legends ``RR-SUS CoMP" and ``RR-SUS non-CoMP" respectively stand
for RR-SUS under CoMP and non-CoMP systems, all with full CSI.}}
\label{F:cdfperfect-0dB}
\end{figure*}

Figure~\ref{F:cdfperfect-10dB} shows the CDF of user data rate
achieved by LargeUS and SUS with $N_t=2, 4$ and $\text{SNR}_0 =
10$~dB, where 3~dB and 10~dB cell-edge regions are considered,
respectively. Compared to non-CoMP systems, CoMP transmission
provides an evident performance gain as expected. In CoMP systems,
the performance gap between RR-LargeUS and RR-SUS decreases when
$N_t$ increases from 2 to 4. This is because more antennas can
improve the accuracy of the approximation used in deriving the
scheduling metrics as shown in Fig.~\ref{F:tightness}(b). For large
cell-edge regions, the approximation is accurate, also as shown in
Fig.~\ref{F:tightness}(b). Consequently, we can see from
Fig.~\ref{F:cdfperfect-10dB}(b) that RR-LargeUS performs close to
RR-SUS. For small cell-edge regions, the channels become not so
asymmetric that the approximation is not very accurate. Although
this will lead to performance degradation for RR-LargeUS,
Fig.~\ref{F:cdfperfect-10dB}(a) shows that the performance loss
compared to RR-SUS is small when $N_t = 4$. Similar results at
$\text{SNR}_0 = 0$~dB can be observed in
Fig.~\ref{F:cdfperfect-0dB}, but the performance gain of CoMP over
non-CoMP reduces because the systems become noise-limited.

\begin{figure*}[!t]
\centering \subfigure[3~dB cell-edge
region]{\includegraphics[width=0.46\textwidth]{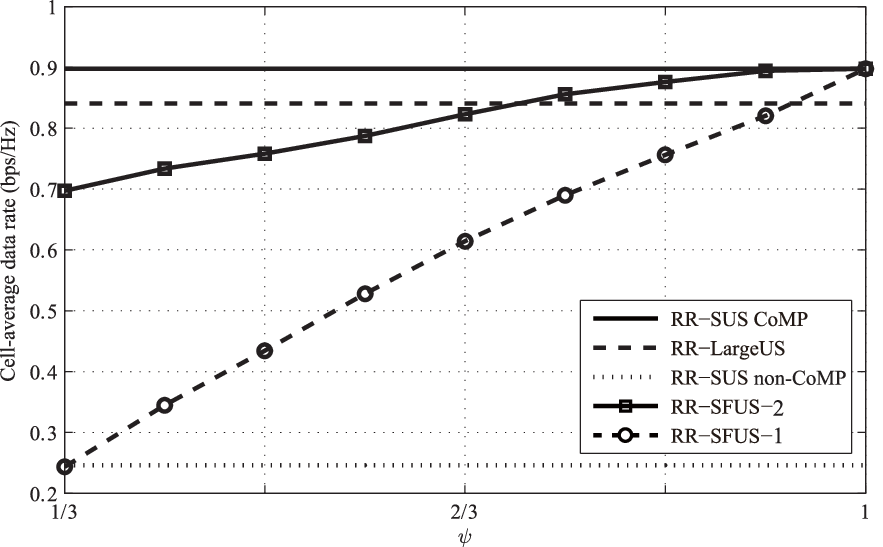}
\label{F:SFUS3dB}} \subfigure[10~dB cell-edge
region]{\includegraphics[width=0.46\textwidth]{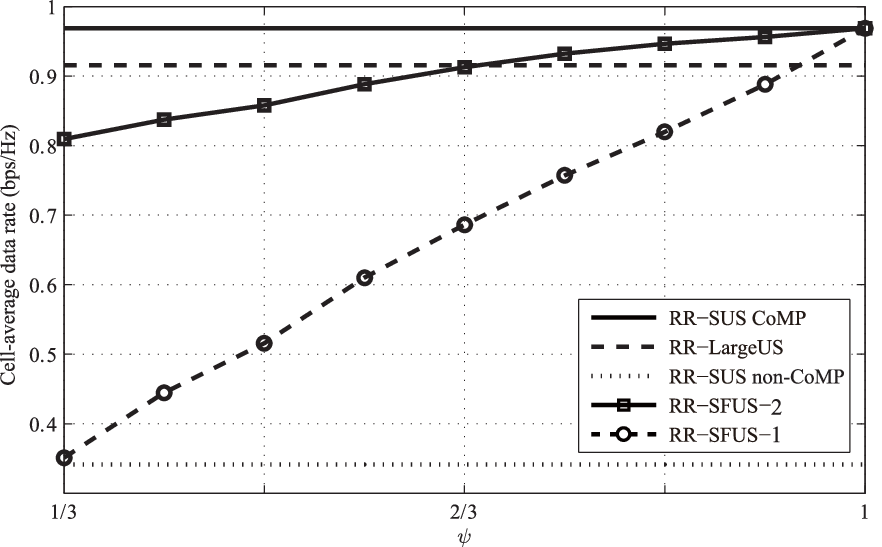}
\label{F:SFUS10dB}}
\caption{Cell-average data rate of RR-SFUS as a function of $\psi$
for various cell-edge regions with $N_t = 4$ and $\text{SNR}_0 =
10$~dB. RR-SFUS schedules users using SUS with full local CSI and
some full cross CSI with large receive power of all candidate users,
where the thresholds $\epsilon$ are properly chosen for different
$\psi$. After scheduling, the ZFBF precoder is computed based on
either the same incomplete CSI as scheduling (denoted by
``RR-SFUS-1'' in the legend) or full global channels of the selected
users (denoted by ``RR-SFUS-2'').} \label{F:SFUS}
\end{figure*}

\begin{figure*}[!t]
\centering \subfigure[3~dB cell-edge
region]{\includegraphics[width=0.46\textwidth]{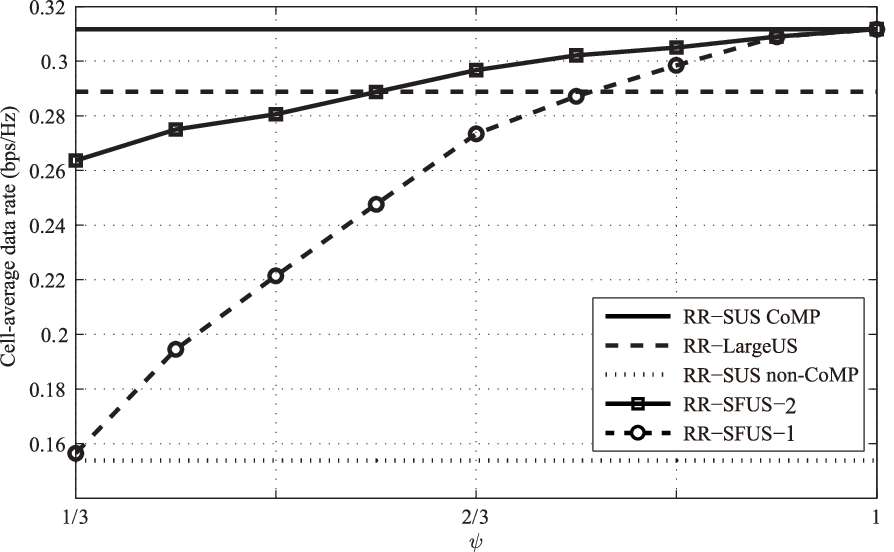}
\label{F:SFUS3dB-2}} \subfigure[10~dB cell-edge
region]{\includegraphics[width=0.46\textwidth]{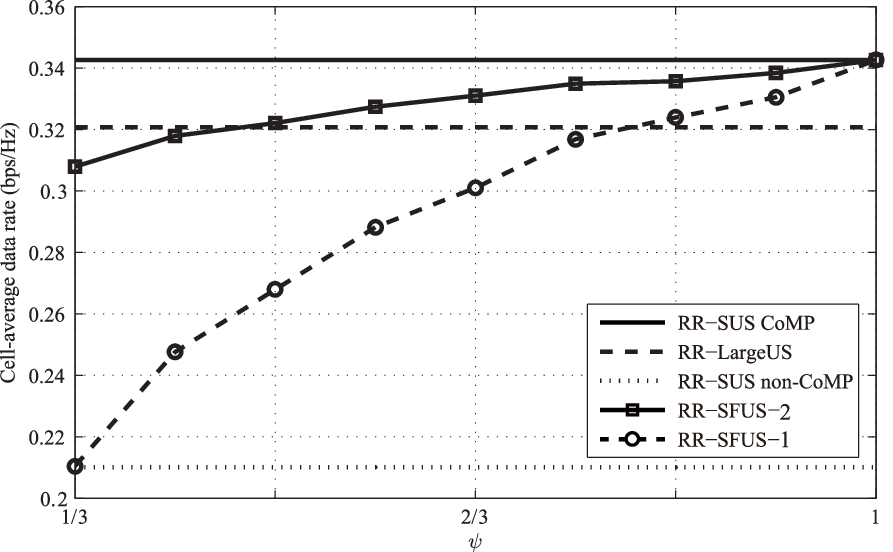}
\label{F:SFUS10dB-2}}
\caption{{Cell-average data rate of RR-SFUS as a function of $\psi$
for various cell-edge regions with $N_t = 4$ and $\text{SNR}_0 =
0$~dB. RR-SFUS schedules users using SUS with full local CSI and
some full cross CSI with large receive power of all candidate users,
where the thresholds $\epsilon$ are properly chosen for different
$\psi$. After scheduling, the ZFBF precoder is computed based on
either the same incomplete CSI as scheduling (denoted by
``RR-SFUS-1'' in the legend) or full global channels of the selected
users (denoted by ``RR-SFUS-2'').}} \label{F:SFUS-2}
\end{figure*}

Finally, we compare the cell-average data rate of the low-overhead
RR-SFUS and the relevant schedulers with $N_t=4$ in
Fig.~\ref{F:SFUS} and Fig.~\ref{F:SFUS-2}, where $\text{SNR}_0$ is
set to 10~dB and 0~dB, respectively. Based on the idea of reducing
feedback overhead for frequency division duplex (FDD) systems
proposed in~\cite{Papadogiannis2011}, in TDD systems RR-SFUS can
operate as follows.
\begin{enumerate}
    \item[(1)] \emph{Channel acquisition:} The coordinated BSs need to know at least full local
channels of all users. In addition, the full cross channels with
large receive power also need to be provided to BSs. Let $\psi$
denote the ratio of the number of selected channel coefficients to
global channels. Then, we have $\frac{1}{M}\leq\psi\leq 1$. The
value of $\psi$ depends on a predetermined receive power
threshold~\cite{Papadogiannis2011}, by adjusting which we consider
different $\psi$ in Fig.~\ref{F:SFUS} and Fig.~\ref{F:SFUS-2}.
\item[(2)] \emph{Scheduling:} With the incomplete CSI (the cross channels with lower receive power are set to
zeros), RR-SFUS schedules users by using the method of SUS in a RR
fashion. Note that a scheduler was proposed for the selective
feedback strategy in~\cite{Papadogiannis2011} but aimed at reducing
backhauling loads and hence is not suitable for our considered
scenario.
\item[(3)] \emph{Precoding:} Based on the types of CSI used for precoding, two RR-SFUS strategies
are considered as follows:
\begin{itemize}
  \item \emph{RR-SFUS-1:} The same incomplete CSI for
  scheduling, including full local CSI and
some full cross CSI, is used for ZFBF precoding. Therefore, in this
case both scheduling and precoding are performed in one stage (denoted by RR-SFUS-1). Since the precoder is computed based on
incomplete global channels, RR-SFUS-1 cannot thoroughly eliminate
the inter-cell interference.
\item \emph{RR-SFUS-2:} As an alternative, RR-SFUS can also be applied for a two-stage
transmission strategy (denoted by RR-SFUS-2), where in the first
stage, RR-SFUS is performed based on the incomplete global channels
and in the second stage, ZFBF precoder is computed with full CSI of
the selected users.
\end{itemize}
\end{enumerate}

We can see from Fig.~\ref{F:SFUS} that RR-SFUS-1 suffers significant
performance loss compared to RR-SFUS-2. To achieve the same
performance as RR-LargeUS, around 90\% of the channel coefficients
need to be provided to the BSs for all considered cell-edge regions.
The performance gap between RR-SFUS-2 and RR-LargeUS when $\psi=1/3$
reflects the importance of using large-scale gains of cross
channels, which are effectively used in the scheduling metrics of
RR-LargeUS but not in RR-SFUS-2. In this case, RR-SFUS-2 exploits
only local channels and regards cross channels as zeros, which is
equivalent to assume that the users in different cells are spatially
separated. Yet, this is far from the reality as shown in
Fig.~\ref{F:tightness}(a). To recover the performance gap, more
channel coefficients need to be provided for RR-SFUS-2. The amount
increases with the shrinking of cell-edge region, e.g., 2/3 channel
coefficients need to be provided for a 10~dB cell-edge region.
Similar results have been observed for the cell-edge data rate,
which are not shown due to space limitations. Fig.~\ref{F:SFUS-2}
considers a noise-limited scenario with $\text{SNR}_0 = 0$~dB. It
can be observed that almost half of channel coefficients are
required by RR-SFUS-2 to achieve the performance of the proposed
LargeUS for a 10~dB cell-edge region, and even more are required for
a 3~dB cell-edge~region.

It is well understood that RR-SFUS can reduce the feedback overhead
of FDD systems by not feeding back the weak cross channels. In the
considered TDD systems, according to the principle of FDD systems,
only local channel and some strong cross channels of each user need
to be estimated, while some weak cross channels can be simply set to
zeros. However, uplink training design for estimating a part of
channel coefficients of global channels has not been well addressed
so far. Thereby we cannot exactly measure the overhead of RR-SFUS-2.
In general, nevertheless, the training overhead increases with a
growing number of channel coefficients to be estimated, which is
determined by $\psi$. This suggests that the training overhead is
maximal when $\psi=1$, i.e., estimating full global channels of all
candidate users, which is $MK$-dimensional resources as explained in
Section \ref{S:Trainingoverhead}. The lower bound of the training
overhead can be obtained when $\psi=1/3$. In this case, the BSs only
estimate local channels of all candidate users for scheduling in the
first stage. The training overhead is the same as that in non-CoMP
systems, which needs $K$-dimensional resources. In the second stage,
$L$-dimensional resources are used to estimate full global channels
of $L$ scheduled users for precoding. Therefore, the lower bound of
training overhead is $(K+L)$-dimensional resources in total. By
contrast, RR-LargeUS only employs large-scale fading gains for
scheduling and requires much less overhead.

\section{Conclusions}
We have studied low-overhead user scheduling for CoMP systems. We
showed that the orthogonality of users' channels can be judged by
their large-scale fading gains when the channels are asymmetric.
Based on this observation, we proposed new scheduling metrics only
depending on average channel gains, with which a low-overhead user
scheduler was developed. Simulation results showed that the proposed
scheduler with large-scale fading gains performs close to the
semi-orthogonal scheduler with full channel information even when
the users are located in cell-edge regions, and requires much less
training overhead than the selective feedback strategy to achieve
the same performance.

\appendices
\section{} \label{A:Jointpdf}
Here we derive the PDF of $\cos^2\theta$ considering $\mathbf{h}_{i}
\sim \mathcal{CN}(\mathbf{0}, \mathbf{R}_i)$ with $\mathbf{R}_i =
\mathrm{diag}\{\alpha_{i1}{\mathbf{I}}, \dots,
\alpha_{iM}{\mathbf{I}}\}$ for $i=1, 2$. When
$\mathbf{R}_i$ is a scaled identity matrix
, $\cos^2\theta$ has been shown to follow a beta distribution with
parameters $1$ and $N-1$~\cite{Yoo06-SUS}, where $N=MN_t$. Since
CoMP channels are asymmetric, $\mathbf{R}_i$ is no longer a scaled
identity matrix.

Note that $\cos^2\theta =
\frac{|\mathbf{h}_2\mathbf{h}_1^H|^2}{\|\mathbf{h}_2\|^2\|\mathbf{h}_1\|^2}
  = \frac{|\mathbf{h}_2\mathbf{v}_1^H|^2}{\|\mathbf{h}_2\|^2}$,
where $\mathbf{v}_1 = \mathbf{h}_1/\|\mathbf{h}_1\|$. Define $q_n =
|\mathbf{h}_2\mathbf{v}_n^H|^2$ and $\mathbf{q} = [q_1,\dots,q_N]$,
where $\mathbf{V} = [\mathbf{v}_1^T, \dots,\mathbf{v}_{N}^T]^T$ is a
standard orthogonal basis generated from $\mathbf{v}_1$. Then
$\cos^2\theta = \frac{q_1}{\sum_{n=1}^{N}{q_n}}$ and we can obtain
its PDF if the joint PDF of $\mathbf{q}$,
$f_\mathbf{q}(\mathbf{x})$, is available by
\begin{align} \label{E:pdf_cos}
 &f_{\cos^2\theta}(x) = \int_{0}^\infty \cdots \int_{0}^{\infty} \frac{\sum_{n=2}^Ny_n}
  {(1-x)^2}\nonumber\\
  &\quad \quad \ \ \ \ \ \cdot f_{\mathbf{q}}\left(\frac{x\sum_{n=2}^Ny_n}{1-x},y_2,\dots,y_N\right)dy_2\dots
  dy_N.
\end{align}

To derive $f_\mathbf{q}(\mathbf{x})$, we first obtain the PDF of
$\mathbf{v}_1$, then derive conditional joint PDF of $\mathbf{q}$
given~$\mathbf{v}_1$.

Define $\mathbf{h}_1 =
[\sqrt{\xi}_1e^{j\phi_1},\dots,\sqrt{\xi}_Ne^{j\phi_N}]$ and $\eta =
\|\mathbf{h}_1\|^2=\sum_{n=1}^N\xi_n$. Then $\mathbf{v}_1$ can be
expressed as $\mathbf{v}_1 =
[\sqrt{\delta_1}e^{j\phi_1},\dots,\sqrt{\delta_N}e^{j\phi_N}]$ with
$\delta_n = \xi_n/\eta$, where $0\leq\delta_n\leq 1$, $0\leq
\phi_n\leq 2\pi$, $n = {1,\dots,N}$, and $\delta_N =
1-\sum_{n=0}^{N-1}\delta_n$. The joint PDF of $\eta$, $\pmb{\xi}$
and $\pmb{\phi}$,
$f_{\eta,\pmb{\xi},\pmb{\phi}}(x,\mathbf{y},\mathbf{z})$, is given
in~\cite{Hammarwall08}, where $\pmb{\xi} = [\xi_1,\dots,\xi_{N-1}]$
and $\pmb{\phi}= [\phi_1,\dots,\phi_N]$, from which we can obtain
the joint PDF of $\pmb{\delta}$ and $\pmb{\phi}$ as
\begin{equation} \label{E:deltaphi}
  f_{\pmb{\delta},\pmb{\phi}}(\mathbf{y},\mathbf{z})=\int_0^\infty x^{N-1}f_{\eta,\pmb{\xi},\pmb{\phi}}(x,\mathbf{y},\mathbf{z})dx,
\end{equation}
where $\pmb{\delta} = [\delta_1,\dots,\delta_{N-1}]$ and $x^{N-1}$
is the Jacobian determinant.

Given $\pmb{\delta}$ and $\pmb{\phi}$ (i.e., given $\mathbf{v}_1$),
it is not hard to find that the vector
$[\mathbf{h}_2\mathbf{v}_1^H,\dots,\mathbf{h}_2\mathbf{v}_N^H]$
follows the joint complex Gaussian distribution
$\mathcal{CN}(\mathbf{0}, \mathbf{V}\mathbf{R}_2\mathbf{V}^H)$. Note
that $q_n = |\mathbf{h}_2\mathbf{v}_n^H|^2$. Then following the work
in~\cite{Krishnamoorthy51}, we can get the conditional joint PDF of
$\mathbf{q}$ given $\pmb{\delta}$ and $\pmb{\phi}$ as
\begin{align} \label{E:Cpdf_y}
  &f_{\mathbf{q}|\pmb{\delta},\pmb{\phi}}(\mathbf{q})= \sum_{r=0}^\infty\frac{(1/2)_r}{r!}
  \frac{e^{-\sum_{n=1}^N\frac{q_n}{a_{nn}}}}{\prod_{n=1}^Na_{nn}} \Bigg(1-\sum_{n_1,\dots,n_N=0}^{2}\nonumber\\
  & \quad \quad \quad \left.C_{n_1,\dots,n_N}\left[\frac{L(\frac{q_1}{a_{11}})}{a_{11}}\right]^{n_1}\dots \left[\frac{L(\frac{q_N}{a_{NN}})}{a_{NN}}\right]^{n_N}\right)^r,
\end{align}
where the function $L(x)$ satisfies $[L(x)]^m[L(x)]^n=[L(x)]^{m+n}$
and $[L(x)]^m \equiv L_m(x)$, $L_m(x)$ is the Laguerre polynomials
of degree $m$~\cite[(8.970)]{Gradshteyn00}, the operator $(x)_{r} =
x(x+1)\dots(x+r-1)$, and $C_{n_1,\dots,n_N}=(n_1!\dots
n_N!)^{-1}\frac{\partial^{n_1+\dots+n_N}g(\pmb{\beta})}{\partial
\beta_1^{n_1}\dots\partial
    \beta_N^{n_N}}$ is the Taylor expansion
coefficient of $g(\pmb{\beta}) = \big|\begin{smallmatrix}
\tilde{\mathbf{A}} & \tilde{\mathbf{B}}\\ -\tilde{\mathbf{B}} &
\tilde{\mathbf{A}}
\end{smallmatrix}\big|$ around the point $\pmb{\beta} = [\beta_1, \dots, \beta_N] =
\mathbf{0}$.
%
Let $[\mathbf{X}]_{i,j}$ denote the element at $i$th row and $j$th
column of matrix $\mathbf{X}$. Then $\tilde{\mathbf{A}}$ and
$\tilde{\mathbf{B}}$ are defined as $[\tilde{\mathbf{A}}]_{i,i} =1$,
$[\tilde{\mathbf{A}}]_{i,j} =a_{ij}\beta_j$,
$[\tilde{\mathbf{B}}]_{i,i} =0$, and $[\tilde{\mathbf{B}}]_{i,j}
=b_{ij}\beta_j$ for $i,j\in\{1,\dots,N\}$, where
$a_{ij}=\Re\{[\mathbf{V}\mathbf{R}_2\mathbf{V}^H]_{i,j}\}$ and
$b_{ij}=\Im\{[\mathbf{V}\mathbf{R}_2\mathbf{V}^H]_{i,j}\}$.

Based on (\ref{E:deltaphi}) and (\ref{E:Cpdf_y}), we obtain the
joint PDF of $\mathbf{q}$ as
\begin{align} \label{E:pdf_y}
  f_{\mathbf{q}}(\mathbf{x})&= \int_{0\leq z_1,\dots,z_N\leq 2\pi} \int_{y_1+\dots+y_{N-1}\leq 1}
  f_{\mathbf{q}|\pmb{\delta},\pmb{\phi}}(\mathbf{x})\nonumber\\
  &\qquad \qquad \qquad \qquad \qquad \qquad \cdot f_{\pmb{\delta},\pmb{\phi}}(\mathbf{y},\mathbf{z})d\mathbf{y}d\mathbf{z}.
\end{align}

For a special case of two-BS cooperation each with one antenna, it
is not hard to get a single integral representation of
$f_{\cos^2\theta}(x)$ as (\ref{E:pdf_cos_2}).

\end{document}